\documentclass[]{aa}

\usepackage{graphicx}
\usepackage{txfonts}
\begin{document}
   \title{Multi-frequency investigation of the parsec- and kilo-parsec-scale radio
   structures in high-redshift quasar PKS~1402$+$044}

%  \subtitle{}

   \author{J.~Yang\inst{1,2}
          \and L.I.~Gurvits\inst{3}
          \and A.P.~Lobanov\inst{4}
          \and S.~Frey\inst{5,6}
          \and X.-Y. Hong \inst{1}
          }

   \offprints{J. Yang}

   \institute{Shanghai Astronomical Observatory, Chinese Academy of Sciences\\
              80 Nandan Road, 200030 Shanghai, P.R.~China, \email{junyang@shao.ac.cn}
         \and Graduate University of the Chinese Academy of Sciences, Beijing, P.R.~China
         \and Joint Institute for VLBI in Europe, P.O.~Box~2, 7990~AA Dwingeloo, The Netherlands\\
             \email{lgurvits@jive.nl}
         \and Max-Planck-Institut f\"ur Radioastronomie, Auf~dem H\"ugel 69, D-53121 Bonn, Germany \\
             \email{alobanov@mpifr-bonn.mpg.de}
         \and F\"OMI Satellite Geodetic Observatory, P.O.~Box~585, H-1592 Budapest, Hungary \\
             \email{frey@sgo.fomi.hu}
         \and MTA Research Group for Physical Geodesy and Geodynamics, P.O.~Box~91, H-1521 Budapest, Hungary
             }

   \date{Received ... , 2008; accepted ..., 2008}

% \abstract{}{}{}{}{}
% 5 {} token are mandatory

  \abstract
   {}
   {We investigate the frequency-dependent radio properties of the jet
of the luminous high-redshift ($z=3.2$) radio quasar
PKS~1402$+$044 (J1405$+$0415) by means of radio interferometric
observations.}
   {The observational data were obtained with the VLBI Space
Observatory Programme (VSOP) at 1.6 and 5~GHz, supplemented by
other multi-frequency observations with the Very Long Baseline
Array (VLBA; 2.3, 8.4, and 15~GHz) and the Very Large Array (VLA;
1.4, 5, 15, and 43~GHz). The observations span a period of 7
years.}
   {We find that the luminous high-redshift quasar
PKS 1402$+$044 has a pronounced ``core-jet'' morphology from the
parsec to the kilo-parsec scales. The jet shows a steeper spectral
index and lower brightness temperature with increasing distance
from the jet core. The variation of brightness temperature agrees
well with the shock-in-jet model. Assuming that the jet is
collimated by the ambient magnetic field, we estimate the mass of
the central object as $\sim10^9M_\odot$. The upper limit of the
jet proper motion of PKS~1402$+$044 is 0.03 mas~yr$^{-1}$
($\sim3c$) in the east-west direction.}
{}
   \keywords{galaxies: individual: PKS~1402$+$044
          -- galaxies: active
          -- quasar: general
          -- galaxies: jets
          -- radio continuum: galaxies}
   \titlerunning{Dual-frequency VSOP imaging of the quasar PKS~1402$+$044}
   \authorrunning{J. Yang et al.}
   \maketitle

%
%________________________________________________________________

\section{Introduction}
Very Long Baseline Interferometry (VLBI) studies of high-redshift
quasars at a given observing frequency $\nu_\mathrm{obs}$ can
facilitate comparison of their structural properties with those of
their lower-redshift counterparts at a higher frequency,
$\nu_\mathrm{obs}=\nu_\mathrm{em}/(1+z)$, where $\nu_\mathrm{em}$
is the emitted (rest-frame) frequency and $z$ the redshift of a
distant quasar. High-redsfhit quasars provide indispensable input
in all kinds of studies of the redshift-dependent properties of
extragalactic objects, such as the apparent ``angular
size~--~redshift'' (``$\theta$~--~$z$'', e.g. Gurvits et al.
\cite{gur99}) and ``proper motion~--~redshift'' (``$\mu$~--~$z$'',
e.g. Kellermann et al. \cite{kel99}) relations.

A statistical study of 151 quasars imaged with VLBI at 5~GHz (Frey
et al. \cite{fre97}) demonstrated an overall trend of a decreasing
jet-to-core flux density ratio with increasing redshift, which
could be explained by the difference in spectral indices of cores
and jets. Furthermore, a majority of radio QSOs at $z>4$ seemed to
be even more compact than expected from the direct extrapolation
of the properties of quasars at $z<4$ (Paragi et al.
\cite{par99}).

A number of morphological studies of high-redshift objects have
been made with the Japanese-led Space VLBI mission VSOP (VLBI
Space Observatory Programme). Observations with the VSOP utilised
an array consisting of a group of Earth-based radio telescopes and
an 8-m space-borne antenna on board the satellite HALCA
(Hirabayashi et al. \cite{hir98}). The orbiting antenna with an
apogee of $\sim21\,000$~km and perigee of $\sim560$~km provided
milli-arcsecond (mas) and sub-mas resolution at the frequencies of
1.6 and 5~GHz. VSOP observations at 1.6~GHz provided roughly the
same angular resolution as Earth-based VLBI observations at 5~GHz.
Thus, dual-frequency VSOP observations made it possible to map the
distribution of spectral index across the source structure (e.g.
Lobanov et al. \cite{lob06}) and to study frequency-dependent
structural properties.

PKS~1402$+$044 (J1405$+$0415) is a flat-spectrum radio source from
the Parkes 2.7-GHz Survey. In optics, it is a 19.6-magnitude ($g$
filter) stellar object at a redshift of $z=3.215$. It is a weak
X-ray source with count rates of
$(5.6\pm1.2)\times10^{-3}$~ct~s$^{-1}$, over the band 0.2~--~4~keV
in the Einstein IPC survey database (Thompson et al. \cite{tho98})
and $(1.3\pm0.2)\times10^{-2}$~ct~s$^{-1}$ over the band
0.1~--~2.4~keV in the ROSAT observation (Siebert et al.
\cite{sie98}). The Multi-Element Radio Linked Interferometer
Network (MERLIN) observations of PKS~1402$+$044 made at 1.6~GHz
indicates that there is a secondary component at a separation of
$0\farcs8$ at a position angle of $-123^{\circ}$ and a faint
extended emission at a distance of $3\farcs3$ at a position angle
of $-106^\circ$. VLBI observations at 5~GHz (Gurvits et al.
\cite{gur92}) found that the main component consists of a compact
core and a resolved jet extending up to $\sim18$~mas to the west.

The quasar PKS~1402$+$044 represents a relatively rare case of a
strong radio source at $z>3$ and therefore a potentially rewarding
target for structural studies within a broad range of angular
scales. VSOP observations with their record-high angular
resolution at 1.6 and 5~GHz facilitate direct comparison of
structural properties of PKS~1402$+$044 with its more abundant
strong radio quasars at lower redshifts at the same emitting
frequency.

In this paper, we present VSOP images at 1.6 and 5~GHz, a 15-GHz
Very Long Baseline Array (VLBA) image, and Very Large Array (VLA)
images at 1.4, 5, 15, and 43~GHz of the quasar PKS~1402$+$044;
discuss its spectral properties and brightness temperature
variation along the jet; and determine some physical parameters of
the core and the jet. Throughout the paper, we define the spectral
index $\alpha$ as $S_\nu\propto\nu^{\alpha}$ and adopt the
$\Lambda$CDM cosmological model (Riess et al. \cite{rie04}) with
$H_0=75$~km~s$^{-1}$ Mpc$^{-1}$, $\Omega_\mathrm{m}=0.3$, and
$\Omega_\Lambda=0.7$. In the latter model, the linear scale factor
for PKS~1402$+$044 is $\sim7$~pc~mas$^{-1}$.
%__________________________________________________________________

\section{Observations and data reduction}
\subsection{VSOP experiment}
\begin{figure}
\centering
\includegraphics[width=0.45\textwidth]{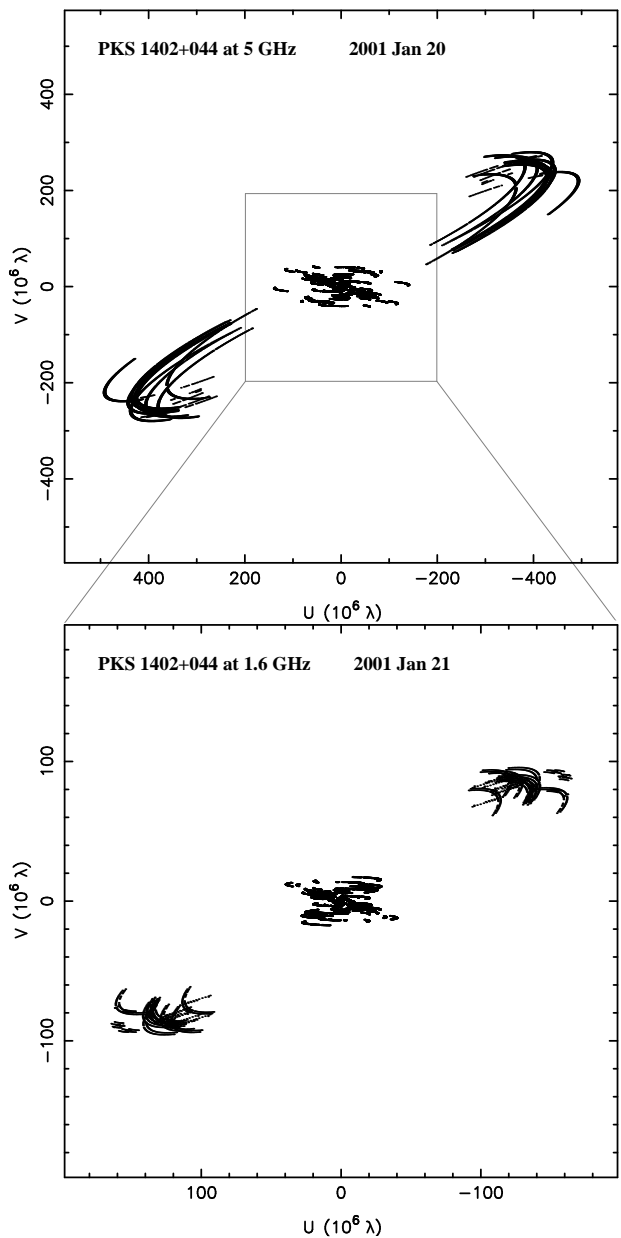} \\
\caption{The effective ($u$,$v$) coverage of the VSOP observations
of PKS 1402$+$044 at 1.6 GHz (bottom) and 5 GHz (top). At each
frequency, the inner tracks correspond to the ground-ground
baselines, and the outer tracks denote the space-ground baselines.
The rectangle in the 5-GHz ($u$, $v$) coverage shows the size of
the 1.6-GHz ($u$, $v$) coverage. The space-ground baselines
provide $uv$-ranges roughly 2.5~--~3 times longer than
ground-ground baselines.} \label{fig1}
\end{figure}
Using the space-borne radio telescope HALCA and the VLBA, we
observed the quasar PKS~1402$+$044 in left hand circular
polarization at 5~GHz on 20~January 2001 and at~1.6 GHz on
21~January 2001. The observations lasted for 8~h at~1.6 GHz and
7~h at 5~GHz. The data were recorded using the VLBA tape system
with a 32-MHz bandwidth consisting of 2 intermediate frequency
(IF) bands and 2-bit sampling, corresponding to the data rate of
128~Mbps. Four tracking stations (Goldstone, Robledo, Tidbinbilla,
Green Bank) were used to receive the HALCA downlink data. The
HALCA data were recorded for $\sim$5~h at 1.6~GHz and 4.2~h at
5~GHz. The data were correlated at the VLBA correlator in Socorro
with 128 spectral channels and an integration time of 4.2~s for
the ground-ground baselines, and 2.1~s at 1.6~GHz, 1.0~s at 5~GHz
for the space-ground baselines. In the 1.6-GHz observation, the
Tidbinbilla station lost 55~minutes of space data and Green Bank
lost all the space data (34~minutes) due to a problem with
recording. At 5~GHz, the IF~1 data were lost for 40~minutes due to
a technical malfunction at the Tidbinbilla station. Except for
these problems, fringes were successfully detected on all the
space-ground baselines at all times.

\begin{figure}
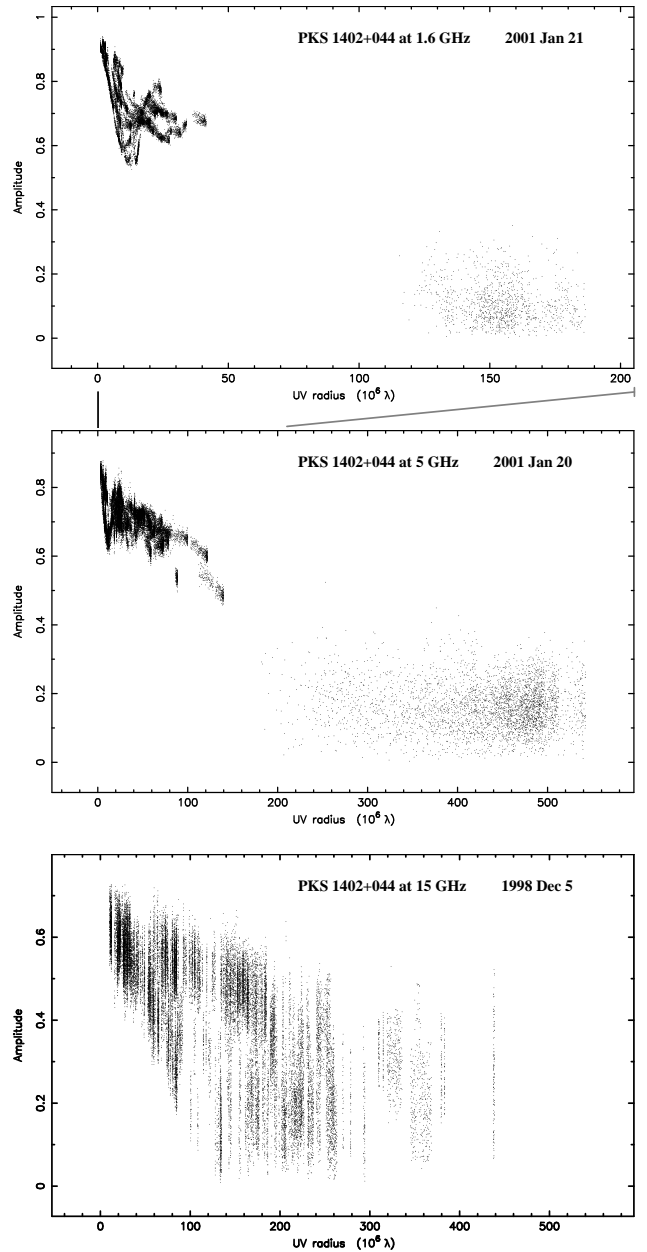

  \centering
  \includegraphics[trim=0 -21 0 14, width=0.45\textwidth,clip]{9846f2a.eps} \\
  \includegraphics[trim=0 -21 0 14, width=0.45\textwidth,clip]{9846f2b.eps} \\
  \includegraphics[trim=0 -21 0 14, width=0.45\textwidth,clip]{9846f2c.eps} \\
\caption{Correlated flux densities of the source PKS 1402$+$044
versus ($u$, $v$) distance at 1.6 GHz (top VSOP data), 5 GHz
(middle VSOP data) and 15 GHz (bottom VLBA data).} \label{fig2}
\end{figure}

\emph{A priori} calibration was applied for both datasets using
the Astronomical Image Processing System (AIPS; Cotton
\cite{cot95}). After correcting the amplitudes in
cross-correlation spectrum using measurements of auto-correlation
spectrum and dispersive delay due to the ionosphere from the maps
of total electron content, we applied \emph{a priori} amplitude
calibration from the antenna gain and system temperature
measurements at the Earth-based telescopes. We used the respective
nominal
values\footnote{http://www.vsop.isas.jaxa.jp/obs/HALCAcal.html}
for HALCA. After inspecting the IF bandpass, the side channels
(1~--~5, 105~--~128) in each IF were deleted because of the lower
amplitude ($<80\%$) than in the centre channels. This reduced the
useful observing bandwidth to 22.75~MHz. Some channels affected by
radio frequency interference were flagged, too. We corrected the
residual delays and rates using a two-step fringe-fitting. We
first fringe-fitted the ground-array data with a solution interval
of 2~minutes. Then we applied the solutions to the data, fixed the
calibration for ground antennas and determined the calibration
solutions of the space antenna using fringe-fitting with a
4-minute interval. After that, we combined all fringe solutions,
applied them to the data, averaged all the channels in each IF,
and split the multi-source data into single-source data sets.
Finally, the data were exported into Difmap (Shepherd et al.
\cite{she94}) and averaged further over 60~s time intervals. The
hybrid imaging and self-calibration were done in Difmap. The
resulting effective ($u$,~$v$) coverages of the VSOP observations
are shown in Fig.~\ref{fig1}. The correlated flux densities as a
function of the projected baseline length are displayed in
Fig.~\ref{fig2}, top and middle.

\subsection{VLBA and VLA data}
The 15-GHz VLBA data presented in this paper are from the
VLBA-VSOP support survey by Gurvits et al. (in preparation). The
observations were conducted on 5~December 1998 with left-hand
circular polarisation, 64-MHz bandwidth and $\sim$50-minute
on-source observing time. We also used the 2.3/8.6~GHz visibility
data provided by US Naval Observatory (USNO) Radio Reference Frame
Image Database (RRFID)\footnote{http://rorf.usno.navy.mil/RRFID}.
Those observations used 10 VLBA antennas and some additional
geodetic antennas. All the VLA data used in this paper were
obtained from the NRAO Data
Archive\footnote{http://archive.nrao.edu/archive/e2earchive.jsp}.
The basic parameters of the VLA observations are summarised in
Table~\ref{tab1}. The columns give (1) frequency in GHz, (2)
program ID, (3) date as dd/mm/yy, (4) array configuration, (5)
antenna numbers, (6) bandwidth in MHz, and (7) total on-source
time in seconds. All the VLA observations used 3C~286 as the prime
flux density calibrator. After \emph{a priori} calibrations in
AIPS, we performed self-calibration, imaging and model-fitting in
Difmap.

\begin{table}[h]
\centering \caption{VLA observations summary.} \label{tab1}
{\footnotesize
\begin{tabular}{ccccccc}
\hline \hline
$\nu_\mathrm{obs}$
     & Program  & Date     & Conf.    &$N_\mathrm{ant}$ & BW  & TOS  \\
(GHz)&          & dd/mm/yy &          &         &(MHz) & (s)  \\
\hline
 1.4 & AH0633   & 11/03/98 & A        & 23       & 100 & 130  \\
 4.8 & AG0670   & 09/10/04 & A        & 26       & 100 & 2010 \\
15.9 & AH0633   & 11/03/98 & A        & 27       & 100 & 170  \\
43.3 & AL0618   & 26/01/04 & BC       & 26       & 100 & 1320 \\
\hline
\end{tabular}
}
\end{table}

%___________________________________________________________________

\section{Results}

\begin{figure*}[]
\centering
\includegraphics[width=\textwidth]{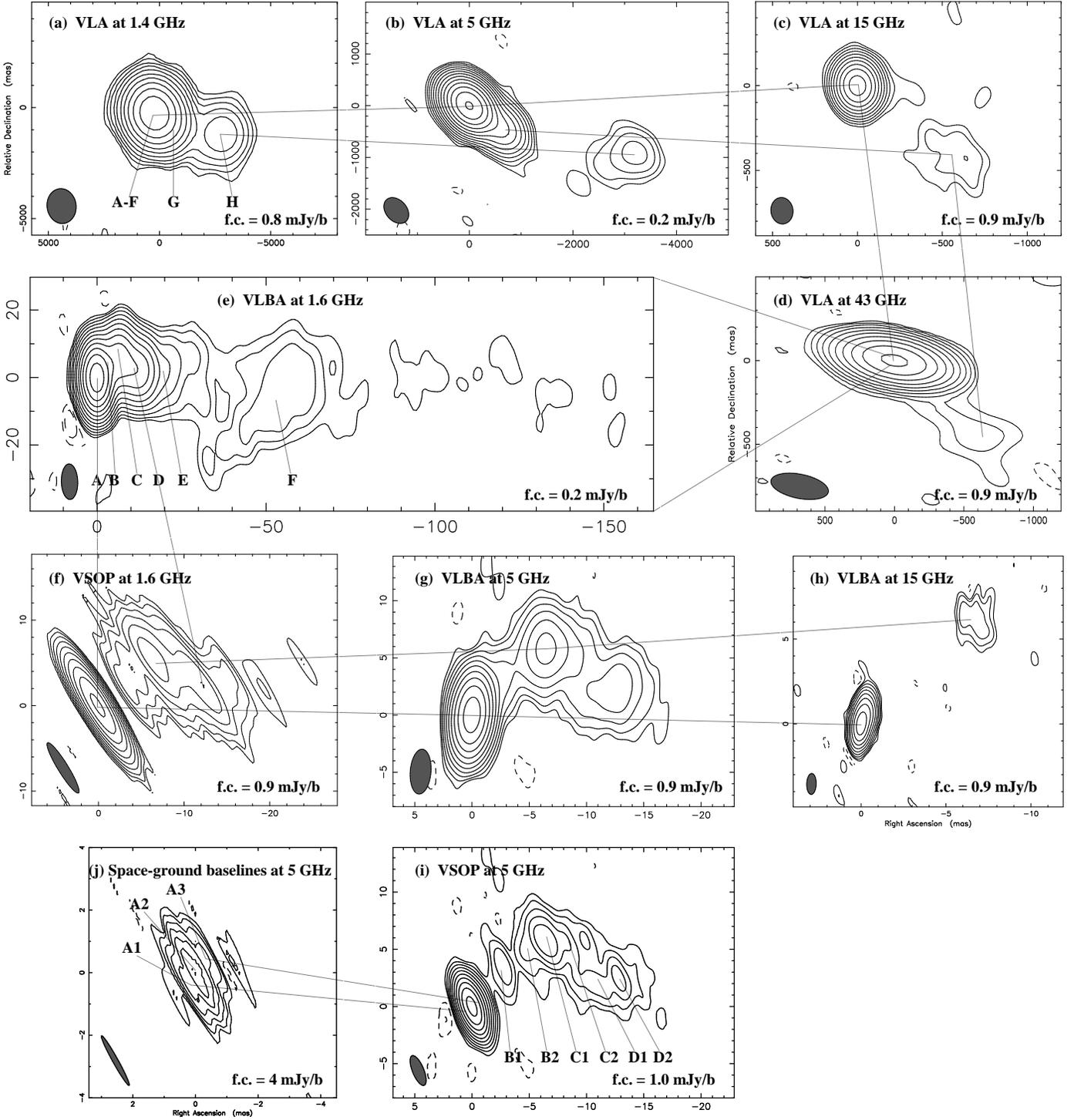}
\caption{VLA and VLBI images of PKS~1402$+$044. The contours are
drawn at $-2$, $-1$, 1, 2, ..., of the respective first contour
(f.c.) levels marked in the bottom of each panel. The latter are
$\sim3\sigma$ of the image thermal noise. The synthesised beams
are plotted in the bottom-left corner of each image. The basic
parameters of each image are listed in Table~\ref{tab2}.}
\label{fig3}
\end{figure*}

Figure~\ref{fig3} shows all the images of PKS 1402$+$044 of the
current study. Their parameters are summarised in
Table~\ref{tab2}. The columns give: (1) frequency in GHz, (2)
array and configuration, (3) weighting (NW: natural, UW: uniform),
(4-5) size of the synthesised beam in mas, (6) position angle of
the major axis in mas, (7) peak brightness in mJy/beam, and (8)
image noise level in mJy/beam (1~$\sigma$). From the final
fringe-fitted VSOP data set we produced two images: (1) an image
with all data included (hereafter VSOP image) and (2) an image
using only the ground VLBA data at each frequency. In the imaging
process, we scaled the gridding weights inversely with the
amplitude errors. The VSOP image fidelity is limited, in
particular, by the completeness of the ($u$, $v$) coverage. In our
case, the latter is essentially one-dimensional (see
Fig.~\ref{fig1}) that leads to a highly elliptical synthesised
beam. However, luckily, the highest angular resolution is achieved
along the position angle of $\sim -60^\circ$, very close to that
of the inner pc-scale jet. Thus, the space-ground baselines
obtained play an important role in imaging the inner pc-scale jet
of PKS 1402$+$044.

\begin{table}[b]
\centering \caption{Parameters of images in
Fig.~\ref{fig3}.}\label{tab2} {\scriptsize
\begin{tabular}{rlrrrrrl}
\hline \hline
$\nu_\mathrm{obs}$
         & Array  &  Wt.  &$b_\mathrm{maj}$
                                  & $b_\mathrm{min}$
                                           & $\theta_\mathrm{pa}$
                                                      &$S_\mathrm{peak}$
                                                               & $\sigma_\mathrm{rms}$ \\
(GHz)    &       &        & (mas) & (mas)  &($^\circ$)& (mJy/b)& (mJy/b)   \\
\hline
 1.4     & VLA:A & NW     & 1580  & 1310   & 7.6      & 862    & 0.3       \\
 4.8     & VLA:A & NW     & 563   & 404    & 41.7     & 919    & 0.07      \\
15.9     & VLA:A & NW     & 156   & 130    & 7.8      & 754    & 0.3       \\
43.3     & VLA:BC& UW     & 419   & 145    & 81.2     & 476    & 0.3       \\
\hline
1.6      & VLBA &   NW    &10.50  &  4.85  &   1.7    &   710  & 0.07      \\
         & VSOP &   UW    & 6.88  &  1.19  &  31.7    &   536  & 0.3       \\
4.8      & VLBA &   NW    & 3.93  &  1.81  &$-$4.5    &   702  & 0.3       \\
         & VSOP &   NW    & 2.79  &  0.99  &  22.3    &   610  & 0.3       \\
         & VSOP &   UW    & 1.81  &  0.17  &  29.1    &   261  & 1.3       \\
15.3     & VLBA &   NW    & 1.24  &  0.56  &$-$1.7    &   370  & 0.3       \\
\hline
\end{tabular}
}
\end{table}

We detect a very weak jet emission extending up to $\sim150$~mas
($\sim$1~kpc) in the high dynamic range ($\sim10\,000$) VLBA image
at 1.6~GHz show Fig.~\ref{fig3}e. It represents a typical core-jet
morphology. We identify a compact core (component~A) and five jet
emission regions (components B~--~F) using circular Gaussian
model-fitting in Difmap. The parameters of the models are listed
in Table~\ref{tab3}. The columns give: (1) component
identification, (2) total flux density of the component, (3 - 4)
radius and position angle of the centre of the component, (5) size
of the fitted circular Gaussian model, (6) the smallest detectable
size, and (7) brightness temperature in the source frame in K. The
error ($1\sigma$) is also listed for all the values. The jet shows
a wide section between 20 and 70~mas (140~--~490~pc projected
distance). The uniformly-weighted VSOP image shown in
Fig.~\ref{fig3}f has a higher angular resolution (at the expense
of considerably higher image noise) and indicates that components
E and F are essentially resolved.

The 5-GHz VLBA image in Fig.~\ref{fig3}g shows a similar structure
to the 1.6-GHz VSOP image and the earlier 5-GHz image by Gurvits
et al. (\cite{gur92}). The naturally-weighted VSOP image at 5~GHz
shows that the jet components at 1.6~GHz are resolved into a few
subcomponents. Here we have differentiated them with postfix
number in the uniformly-weighted space-ground image
(Fig.~\ref{fig3}j). In this image, the jet appears to be heavily
resolved. The core shows a three-component morphology. A weak
component marked as A1 appears at the base of the jet and near the
brightest component A2. The weakness of component A1 may be due to
synchrotron self-absorption considering its high brightness
temperature ($\sim 10^{12}$~K). There are two relatively weak jet
components, B1 and B2, at 1.6 and 5~GHz between the bright
components A and C. At the higher frequency, 15~GHz, both
components are too weak ($<0.9$~mJy/beam) to be detected. Based on
the spectrum at frequencies $\leq5$~GHz, the extrapolated total
flux density of B1$+$B2 is $\sim5$~mJy at 15~GHz. The nondetection
of the two components indicates that they have a steeper spectrum
($\alpha<-0.9$) at frequencies $>5$ GHz.

%____________________________________________________________________

The 1.4-GHz VLA image (Fig.~\ref{fig3}a) has the lowest resolution
and shows that there is a weak ($\sim33$ mJy) component (H) at a
distance of $3\farcs22$ and a position angle $-107\fdg4$ from the
core, besides the main emission region. It agrees well with
earlier MERLIN observations made with the Westerbork Synthesis
Radio Telescope (WSRT; Gurvits et al. \cite{gur92}). The higher
sensitivity (0.07~mJy/beam) VLA observations (Fig.~\ref{fig3}b) at
5~GHz indicate that component H has a weak extension toward east.
The extension is consistent with the hypothesis that component H
belongs to the jet of PKS~1402$+$044. The main emission region can
be approximated by component G and a combination of the inner
components (A~--~F) in the VLA images. Component G is also
detected at 15~GHz in Fig.~\ref{fig3}c and even 43~GHz in
Fig.~\ref{fig3}d. The highest observing frequency corresponds to
the rest-frame emitted frequency of $\sim180$~GHz. Arguably, this
is one of the rare cases of a profound jet emission at millimetre
wavelengths.

\begin{table*}
\centering \caption{ The parameters and the brightness
temperatures of the fitted Gaussian models. }\label{tab3}
\begin{tabular}{rrrrrrl}
\hline\hline
Comp. &$S_\mathrm{int}~~~~$& $r$~~~~~~~~&$\theta_\mathrm{pa}$~~~~~~~~~& $d$~~~~~~~~~~  &$d_\mathrm{lim}$& ~~~~$T_\mathrm{b}$\\
      & (mJy)~~       & (mas)~~     &($^\circ$)~~~~~~~~~~& (mas)~~~~~~    & (mas)   &    ~~~~(K)                \\
\hline
\multicolumn{7}{c}{Naturally weighted VLA image at 1.425 GHz}                                                      \\
A-F   &  $795\pm28$   &    0           & --~--~--        &$83\pm\ \ \ \ 2$&  34    &   $(4.2\pm0.3)\times10^8$      \\
G     &  $131\pm12$   &  $701\pm10$    &  $-124.5\pm0.8$ &$261\pm\ \ 21$ &  84    &   $(6.9\pm1.3)\times10^6$      \\
H     &  $ 33\pm\ \ 8$& $3218\pm83$    &  $-107.4\pm1.5$ & $691\pm165$   & 167    &   $(2.5\pm1.4)\times10^5$      \\
\hline
\multicolumn{7}{c}{Naturally weighted VLA image at 4.860 GHz}                                                      \\
A-F   &  $922\pm33$   &    0           & --~--~--        &$31\pm\ \ \ \ 1$&  11    &   $(1.0\pm0.1)\times10^9$      \\
G     &  $ 47\pm10$   & $ 706\pm25$    &  $-125.3\pm2.0$ & $243\pm\ \ 50$&  51    &   $(8.3\pm3.8)\times10^5$      \\
H     &  $  4\pm\ \ 2$& $3324\pm68$    &  $-106.1\pm1.2$ & $430\pm136$   &  98    &   $(2.5\pm1.8)\times10^4$      \\
\hline
\multicolumn{7}{c}{Naturally weighted VLA image at 14.94 GHz}                                                      \\
A-F   &  $756\pm35$   &    0           & --~--~--        &   $7\pm\ 0.5$   &   4    &   $(5.7\pm0.5)\times10^9$      \\
G     &  $ 15\pm\ \ 5$&  $704\pm48$    &  $-126.6\pm3.9$ & $282\pm\ \ 96$&  25    &   $(6.5\pm5.0)\times10^4$      \\
\hline
\multicolumn{7}{c}{Uniformly weighted VLA image at 43.34 GHz}                                                      \\
A-F   &  $477\pm17$   &    0           & --~--~--        &   $4\pm\ 0.2$ &  12    &   $(3.0\pm0.2)\times10^9$      \\
G     &  $  6\pm\ \ 2$&  $698\pm52$    &  $-127.8\pm4.3$ & $277\pm103$   & 109    &   $(8.6\pm7.4)\times10^3$      \\
\hline
\multicolumn{7}{c}{Naturally weighted VLBA image at 1.646 GHz}                                                     \\
A     &  $713\pm38$   &    0           & --~--~--        &$ 0.77\pm0.03$ & 0.25   &   $(3.8\pm0.4)\times10^{12} $  \\
B     &  $ 35\pm\ \ 8$&  $4.88\pm0.11$ &  $-43.7\pm1.3$  &$ 1.20\pm0.23$ & 0.96   &   $(7.5\pm3.3)\times10^{10} $  \\
C     &  $ 98\pm12$   &  $8.81\pm0.12$ &  $-50.2\pm0.8$  &$ 2.70\pm0.24$ & 0.57   &   $(4.2\pm0.9)\times10^{10} $  \\
D     &  $ 68\pm10$   & $12.32\pm0.23$ &  $-79.8\pm1.1$  &$ 3.92\pm0.47$ & 0.68   &   $(1.4\pm0.4)\times10^{10} $  \\
E     &  $ 13\pm\ \ 6$& $19.32\pm2.07$ &  $-84.6\pm6.1$  &$ 9.14\pm4.14$ & 1.60   &   $(4.7\pm4.6)\times10^8    $  \\
F     &  $ 14\pm\ \ 8$& $55.32\pm5.58$ &  $-94.2\pm5.8$  &$20.82\pm11.2$ & 1.50   &   $(1.0\pm0.9)\times10^8    $  \\
\hline
\multicolumn{7}{c}{Naturally weighted VSOP image at 4.862 GHz}                        \\
A1    &  $134\pm12$   &    0           & --~--~--        & $0.18\pm0.01$ & 0.10   &   $(4.4\pm0.8)\times10^{12} $  \\
A2    &  $454\pm21$   & $ 0.49\pm0.01$ &  $-25.6\pm0.9$  & $0.29\pm0.01$ & 0.05   &   $(5.7\pm0.5)\times10^{12} $  \\
A3    &  $144\pm12$   & $ 1.10\pm0.01$ &  $-29.1\pm0.8$  & $0.41\pm0.03$ & 0.09   &   $(8.9\pm1.4)\times10^{11} $  \\
B1    &  $ 10\pm\ \ 4$& $ 4.58\pm0.11$ &  $-37.4\pm1.4$  & $0.66\pm0.22$ & 0.35   &   $(2.5\pm1.9)\times10^{10} $  \\
B2    &  $  4\pm\ \ 2$& $ 6.65\pm0.15$ &  $-41.7\pm1.3$  & $0.66\pm0.29$ & 0.57   &   $(9.7\pm9.5)\times10^9    $  \\
C1    &  $ 52\pm\ \ 9$& $ 9.12\pm0.13$ &  $-46.4\pm0.8$  & $1.64\pm0.26$ & 0.15   &   $(2.0\pm0.8)\times10^{10} $  \\
C2    &  $ 14\pm\ \ 5$& $10.78\pm0.37$ &  $-54.1\pm2.0$  & $2.21\pm0.73$ & 0.30   &   $(3.1\pm2.3)\times10^9    $  \\
D1    &  $  8\pm\ \ 3$& $10.11\pm0.35$ &  $-73.4\pm2.0$  & $1.84\pm0.69$ & 0.39   &   $(2.6\pm2.2)\times10^9    $  \\
D2    &  $ 26\pm\ \ 8$& $13.18\pm0.38$ &  $-78.0\pm1.6$  & $2.52\pm0.76$ & 0.22   &   $(4.3\pm2.9)\times10^9    $  \\
\hline
\multicolumn{7}{c}{Naturally weighted VLBA image at 15.34 GHz}                                                     \\
A1    &  $102\pm13$   &    0           & --~--~--        & $0.21\pm0.02$ & 0.07   &   $(7.7\pm1.8)\times10^{11} $  \\
A2    &  $287\pm21$   & $0.50\pm0.01 $ &  $-19.5\pm1.3$  & $0.18\pm0.01$ & 0.04   &   $(2.9\pm0.4)\times10^{12} $  \\
A3    &  $207\pm18$   & $1.24\pm0.02 $ &  $-20.9\pm0.9$  & $0.52\pm0.03$ & 0.05   &   $(2.6\pm0.5)\times10^{11} $  \\
C1    &  $ 29\pm\ \ 9$& $9.58\pm0.23 $ &  $-46.0\pm1.4$  & $1.64\pm0.47$ & 0.14   &   $(3.6\pm2.3)\times10^9    $  \\
  \hline
\end{tabular}

\end{table*}

\section{Discussion}
\subsection{Spectral properties of the jet}
The resolution ($6.88\times1.19$~mas) of the Space VLBI image of
PKS~1402$+$044 at 1.6~GHz is close ($3.55\times1.40$~mas) to that
of the ground VLBA image at 5~GHz, enabling extraction of spectral
index information from a combination of the two images. We
restored the 1.6-GHz VSOP image and the 5-GHz VLBA image with a
circular Gaussian beam of 4~mas in diameter. The artificial beam
increases the beam area by a factor of $\sim2$ at 1.6~GHz and
$\sim3$ at 5~GHz compared to the areas of the original synthesised
beams. Both images were aligned at the strongest component. The
shifts in the image centre are less than 0.1 mas ($\ll$~4-mas
resolution). After the alignment, the spectral index was
calculated at all pixels with brightness values higher than 1.8
mJy/beam ($5\sigma$) in the 1.6-GHz image and 1.2~mJy/beam
($5\sigma$) in the 5-GHz image. A possible core shift between the
two frequencies as predicted by Kovalev et al. (\cite{kov08}) does
not exceed 1.5~mas. Thus, it does not affect the large-scale
spectral distribution.

\begin{figure}[ht]
\centering
\includegraphics[height=0.4\textwidth]{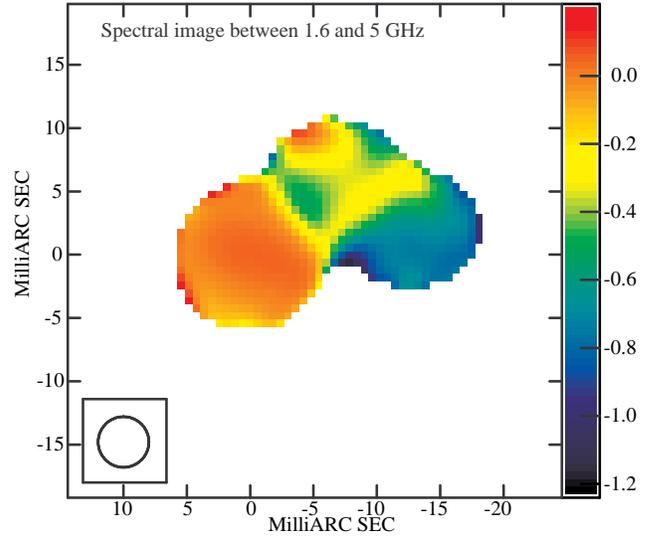} \\
\caption{The spectral index distribution in the jet of the quasar
PKS~1402$+$044.} \label{fig4}
\end{figure}

\begin{figure}[ht]
\centering
\includegraphics[width=0.45\textwidth]{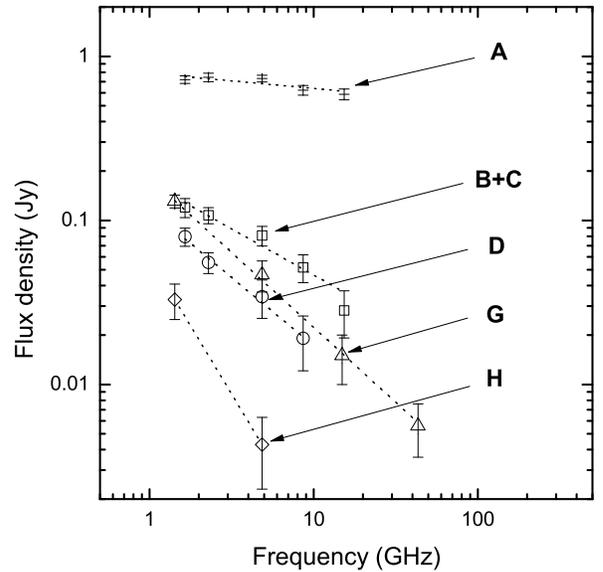}\\
\caption{The component spectra in the jet of the quasar
PKS~1402$+$044. The dotted lines represent the fitted curves with
the simple power-law model $S_\nu=S_0\nu^\alpha$. The estimated
spectral indices are listed in Table~\ref{tab4}.} \label{fig5}
\end{figure}

The final spectral index distribution between 1.6~GHz and 5~GHz is
displayed in Fig.~\ref{fig4}. It shows a smooth distribution of
spectral index on $\sim20$~mas (140~pc) scale. The spectral index
varies from $+$0.1 in the optically thick base region to $-1.0$ in
the optically thin regions on the western side. To further confirm
the variation, we plotted the components spectra in
Fig.~\ref{fig5}. Here we also used the 2.3/8.4-GHz visibility data
from the USNO RRFID database. We fitted the VLBI visibility data
with three components at each frequency (1.6, 2.3, 5, 8.4, and
15~GHz). The spectra of the large-scale components G and H are
also plotted. All spectra can be approximated by a power-law
model, $S_\nu=S_0\nu^{\alpha}$. The spectral indices are listed in
Table~\ref{tab4}. The spectral steepening increases with the
increase in the distance from the core. The spectral difference is
0.47 between components A and B$+$C and reaches 1.57 between
component A and the farthest component H. This spectral index
gradient results in the variation in the flux density ratio of the
components (B+C) over A from $\sim 0.19$ at 1.6 GHz to $\sim 0.05$
at 15~GHz. In a sample of sources at various redshifts, an
increase in redshift is equivalent to the increase in the
intrinsic emitting frequency,
$\nu_\mathrm{em}=\nu_\mathrm{obs}(1+z)$. Thus, a decrease in
jet-to-core flux density ratio with increase in redshift is
expected (Frey et al. \cite{fre97}). The quasar PKS 1402$+$044 has
a spectral difference of $\sim0.6$ at the pc scale, which agrees
well with the prediction $0.55\pm0.43$ by Frey et al.
(\cite{fre97}).

\begin{table}[h]
\centering \caption{Results of the power-law spectral model fits
for each component shown in Fig.~\ref{fig5}.} \label{tab4}
\begin{tabular}{llll}
\hline \hline
Comp.   & ~~$S_0$       &  ~~~~$\alpha$   &  ~~$\chi^2$ \\
        & ~(Jy)         &                 &             \\
\hline
A       & $0.79\pm0.05$ & $-0.09\pm0.04$  & 1.19        \\
B+C     & $0.17\pm0.02$ & $-0.56\pm0.10$  & 0.72        \\
D       & $0.12\pm0.02$ & $-0.83\pm0.19$  & 0.16        \\
G       & $0.18\pm0.09$ & $-0.91\pm0.09$  & 0.08        \\
H       & $0.06\pm0.03$ & $-1.66\pm0.40$  & -~-~-       \\
\hline
\end{tabular}

\end{table}

\subsection{The mass of the central object of PKS 1402$+$044}
The richness of the core-jet morphology in PKS~1402$+$044 makes it
a suitable source for estimating parameters of the central black
hole. The smallest detectable size for a circular Gaussian
component in an image with an rms noise $\sigma_\mathrm{rms}$ is
defined as (Lobanov \cite{lob05}):
\begin{equation}\label{eq:d-lim}
    d_\mathrm{lim} = \frac{2^{2-\beta/2}}{\pi} \left [
   \pi~b_\mathrm{maj}~b_\mathrm{min} \ln 2
   \ln \left( \frac{S_\mathrm{int} / \sigma_\mathrm{rms}}{S_\mathrm{int} / \sigma_\mathrm{rms}-1}\right )\right
   ]^{1/2},
\end{equation}
where $S_\mathrm{int}$ is the integrated flux density of the
component, $b_\mathrm{maj}$ and $b_\mathrm{min}$ are the major and
minor axes of the restoring beam respectively, $\beta=0$ for
uniform weighting and $\beta=2$ for naturally weighting. Based on
the above criterion, except for the size of the combined component
from A to F at 43 GHz, all the sizes estimated from our VLBI and
VLA images in Column~(5) of Table~\ref{tab3} can be taken as the
true sizes of the jet emission regions.

If the component size is related to the physical transverse
dimension of the jet, the mass of the central object can be
estimated assuming that the jet is collimated by the ambient
magnetic field of the host galaxy. The jet components A2 and A3
have the best measurements of the width of the jet close to the
central object, as they are most likely free of the effect of the
adiabatic expanding of the jet and synchrotron self-absorption and
have the higher reliability, $d/\sigma_d>10$, where $d$ and
$\sigma_d$ are the angular size and error of the fitted circular
Gaussian component. For a jet collimated by the ambient magnetic
field $B_\mathrm{ext}$ of the host galaxy, the mass of the central
object $M_\mathrm{BH}$ can be related to the width of the jet
$r_\mathrm{jet}$ (in pc) according to the following relation
(Beskin \cite{bes97}):
\begin{equation}
    M_\mathrm{BH} \simeq r_\mathrm{jet} (B_\mathrm{ext}/B_\mathrm{gr})^{1/2}
    10^{13}M_\odot,
    \label{eq:m-bh}
\end{equation}
where $B_\mathrm{gr}$ is the magnetic field measured at the
Schwarzschild radius $R_\mathrm{gr}$ of the central black hole.
Equation~(\ref{eq:m-bh}) refers to the transverse dimension of the
jet measured at distances comparable to the collimation scale
(typically expected to be located at
$10^2$~--~$10^3~R_\mathrm{gr}$). A typical galactic magnetic field
is $B_\mathrm{ext}\sim10^{-5}$~G (Beck~\cite{bec00}) and one can
expect to have $B_\mathrm{gr}\sim10^4$~G (Field \& Rogers
\cite{fie93}). Based on these parameters, the mass of the central
object is $\sim10^9M_\odot$. The main uncertainty of the mass
estimation arises from the uncertainty in $B_\mathrm{gr}$.
However, the dependence of the mass on the value $B_\mathrm{gr}$
is rather weak, $M_\mathrm{BH}\propto B_\mathrm{gr}^{-0.5}$; with
the magnetic field varying within 4 orders of magnitude, the
estimated central black hole mass varies within two orders of
magnitude. The external magnetic field $B_\mathrm{ext}$ normally
varies within a narrow range around ($10^{-5\pm1}$~G). Thus, the
magnetic field uncertainty should not affect the estimated mass
drastically.

\subsection{Brightness temperature}
Based on the parameters of the Gaussian models listed in
Table~\ref{tab2}, we calculated the brightness temperature of each
component using the following formula (Kellermann \& Owen
\cite{kel88}):
\begin{equation}\label{eq:tb-obs}
    T_\mathrm{b} = 1.22 \times 10^{12} ( 1+z ) \frac{S_\mathrm{int}}{d^2
    \nu^2},
\end{equation}
where $S_\mathrm{int}$ is the integrated flux density in Jy, $d$
the size of a circular Gaussian component in mas, and $\nu$ the
observing frequency in GHz. The estimated brightness temperatures
are listed in the last column of Table~\ref{tab2}. Among these
components, the component A2 has the highest brightness
temperature, $T_\mathrm{B}=(5.7\pm0.5)\times10^{12}$~K, which is
somewhat higher than the inverse Compton limit ($\sim10^{12}$~K;
Kellermann \& Pauliny-Toth \cite{kel69}) but still 10-times lower
than the currently known highest value of $5.8\times10^{13}$~K
found in the BL Lac object AO~0235$+$164 by Frey et al.
(\cite{fre00}). In the equipartition jet model of Blandford \&
K\"onigl (\cite{bla79}), the limiting brightness temperature is
about $3\times10^{11}\delta^{5/6}$ K, where $\delta$ is the
Doppler factor. Comparing the theoretical value with the estimated
brightness temperature of the component A2, we can obtain a
conservative lower limit to the Doppler factor of the inner jet,
$\delta\approx23.7$.

\begin{figure}[]
  \centering
  \includegraphics[width=0.45\textwidth]{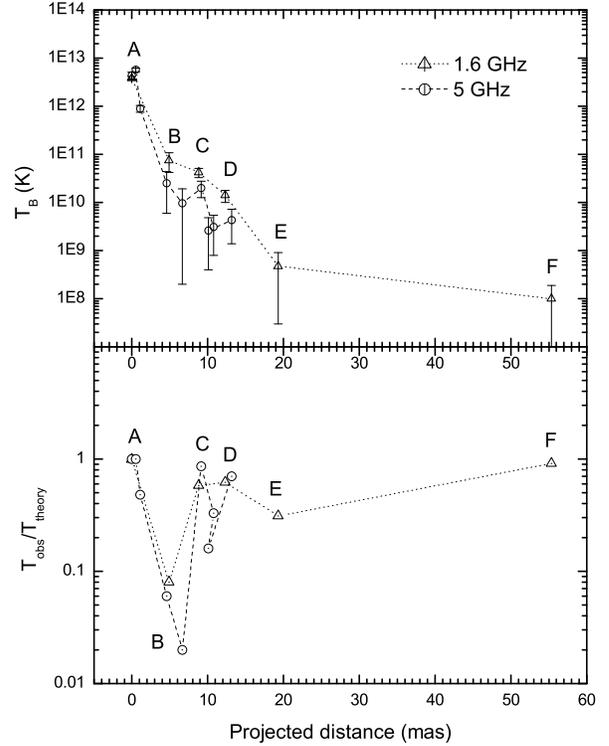}\\
  \caption{
Brightness temperature variations along the jet in PKS 1402$+$044
(top) and comparison with the model predicted value (bottom,
Marscher \cite{mar90}). The component identification is described
in Fig.~\ref{fig3}. The triangles represent the estimated
brightness temperature at 1.6 GHz; the circles represent the
brightness temperature at 5 GHz.}\label{fig6}
\end{figure}

The variation in the observed brightness temperature with
increasing distance from the core is plotted in Fig.~\ref{fig4}.
Following Marscher (\cite{mar90}), we assume that each of the jet
components is an independent plane shock in which the radio
emission is dominated by adiabatic energy losses. The jet plasma
has a power-law energy distribution, $N(E)dE\propto~ E^{-s}dE$.
The magnetic field varies as $B~\propto~l^{-a}$, where $l$ is the
distance from the central object. The Doppler factor is assumed to
vary weakly throughout the jet. Under these assumptions, one can
relate the brightness temperature $T_\mathrm{b,jet}$ of each jet
component to the brightness temperature of the core
$T_\mathrm{b,core}$:
\begin{equation}\label{eq:tb-model}
  T_\mathrm{b,jet} =
  T_\mathrm{b,core}(d_\mathrm{jet}/d_\mathrm{core})^{-\epsilon},
\end{equation}
where $d$ represents the measured size of the core and jet
features and $\epsilon=[2(2s+1)+3a(s+1)]/6$ (Lobanov et al.
\cite{lob01}). We take $s=2.06$ corresponding to the synchrotron
emission with the spectral index of component B+C $\alpha=0.53$,
and $a=1$ corresponding to the transverse orientation of magnetic
field in the jet (Lobanov et al. \cite{lob01}). At each frequency,
we take the brightest component as the core. Comparing with the
measured one, we plotted the ratio of
$T_\mathrm{obs}/T_\mathrm{theory}$ versus the distance. The
largest discrepancies occur with components B at 1.6~GHz, B1, and
B2 at 5~GHz. These components have ratios $d/\sigma_d=5$ at
1.6~GHz and $d/\sigma_d\leq3$ at 5~GHz. The ratios indicate that
the discrepancy may be caused by uncertainties in the size
estimates for strongly resolved components or inhomogeneities in
the plasma over such large emitting regions.

\subsection{Proper motion}
Using an earlier VLBI observation at 5~GHz in 1986 by Gurvits et
al. (\cite{gur92}), we tried to estimate the proper motion in
PKS~1402$+$044 over the time interval of 15~years. We also fitted
the ($u$,$v$) visibility data with 6 Gaussian models and found
that the position offset is within the one fifth of the beam
($10\times2$ mas in P.A.~$-2^\circ$) of the image in 1986 in the
east-west direction where the image has the higher resolution, and
there is no consistent shift. Thus, an upper limit of the apparent
proper motion $\mu$ in the EW direction is 0.03~mas~yr$^{-1}$.
This corresponds to the apparent velocity upper limit of
$\beta_\mathrm{app}=3c$ based on the relation
$\beta_\mathrm{app}=1.58\times10^{-2}\mu D_\mathrm{A}(1+z)$
(Kellermann et al. \cite{kel04}), where the angular size distance
$D_\mathrm{A}$ is measured in Mpc, $\mu$ in mas~yr$^{-1}$ and
$\beta_\mathrm{app}$ in the unit of the speed of light, $c$.

Using the determined lower limit $\delta=23.7$
($\delta\gg\sqrt{\beta_\mathrm{app}^2+1}$) and the following
equations (e.g. Hong et al. \cite{hon08}):
\begin{eqnarray}
  \gamma      &=& \frac{\beta_\mathrm{app}^2 + \delta^2 +1}{2\delta}, \\
  \tan\phi  &=& \frac{2\beta_\mathrm{app}}{\beta_\mathrm{app}^2+\delta^2-1} ,
\end{eqnarray}
a lower limit to the Lorentz factor $\gamma\approx12$ of the jet
and an upper limit to the viewing angle to the line of sight
$\phi\approx1^\circ$ can be determined. All the estimates suggest
that PKS 1402$+$044 is a relativistically beamed radio sources.

\section{Summary}
Based on multi-frequency VLBI (1.6, 2.3, 5, 8.4, and 15 GHz)
observations including dual-frequency (1.6 and 5 GHz) VSOP
observations and VLA (1.4, 5, 15, and 43 GHz) observations of the
high-redshift quasar PKS 1402$+$044, we draw the following
conclusions.
\begin{enumerate}
   \item The quasar PKS 1402$+$044 demonstrates a well-defined core-jet
morphology that can be traced out to $\sim$23 kpc from the source
core.
    \item The radio spectral index distribution and the component spectra prove that
the jet has the steeper spectrum with increasing distance from the
core.
    \item Based on the measurement of the transverse size of the jet, and
assuming that the external magnetic field collimates the jet
model, the mass of the central object is estimated as $\sim 10^9
M_\odot$.
    \item PKS 1402$+$044 has a bright core ($5.7 \times 10^{12}$ K), and the
observed brightness temperature variation is basically consistent
with the shock-in-jet model.
    \item No firm detection of a proper motion in the jet can be made. An upper limit of the apparent proper motion
in the east-west direction is 0.03 mas yr$^{-1}$, corresponding to
the apparent speed of $3~c$.
    \item Based on the lower limit of the Doppler factor $\delta=23.7$
($\delta\gg\sqrt{\beta_\mathrm{app}^2+1}$), we estimate the lower
limit to the Lorentz factor $\gamma=12$ and the upper limit to the
viewing angle of the inner jet to the line of sight as
$\phi=1^\circ$.
\end{enumerate}

\begin{acknowledgements}
We are grateful to Alan~Fey for providing us with the S/X-band
USNO RRFID data, Richard~Schilizzi and Ken~Kellermann for their
assistance at various stages of the project, PI's and the teams of
VLA observations used in this work. The original idea for this
project was conceived in discussions with Ivan~Pauliny-Toth. This
research was partly supported by the Natural Science Foundation of
China (NSFC10473018 and NSFC10333020). Jun~Yang and Xiaoyu~Hong
are grateful to the KNAW~--~CAS grant 07DP010. S\'andor~Frey
acknowledges the OTKA~K72515 and HSO~TP314 grants. We gratefully
acknowledge the VSOP Project, which was led by the Institute of
Space and Astronautical Science (Japan) in cooperation with many
agencies, institutes, and observatories around the world. The
National Radio Astronomy Observatory is a facility of the National
Science Foundation operated under cooperative agreement by
Associated Universities, Inc. This research has made use of NASA's
Astrophysics Data System, NASA/IPAC Extragalactic Database (NED),
and the United States Naval Observatory (USNO) Radio Reference
Frame Image Database (RRFID).
\end{acknowledgements}

\end{document}